\documentclass[twocolumn,letter]{jpsj3}
\usepackage{txfonts}
\usepackage{graphicx}
\usepackage{amsmath}
\usepackage{amssymb}
\usepackage{bm}
\usepackage{subfigure}
\usepackage{color}
\def\figwidth{0.9}

\title{Large-scale Monte Carlo simulation of two-dimensional classical XY model \\
using multiple GPUs} 

\author{Yukihiro Komura\thanks{E-mail: y-komura@phys.se.tmu.ac.jp} 
and Yutaka Okabe\thanks{E-mail: okabe@phys.se.tmu.ac.jp}}

\inst{Department of Physics,  Tokyo Metropolitan University, Hachioji, Tokyo 192-0397, Japan}

\recdate{August 27, 2012}

\abst{
We study the two-dimensional classical XY model by the large-scale 
Monte Carlo simulation of the Swendsen-Wang multi-cluster algorithm 
using multiple GPUs on the open science supercomputer TSUBAME 2.0. 
Simulating systems up to the linear system size $L=65536$, 
we investigate the Kosterlitz-Thouless (KT) transition. 
Using the generalized version of the probability-changing cluster algorithm 
based on the helicity modulus, we locate the KT transition temperature 
in a self-adapted way.  The obtained inverse KT temperature 
$\beta_{\rm KT}$ is 1.11996(6).  
We estimate the exponent to specify the multiplicative 
logarithmic correction, $-2r$, and precisely reproduce 
the theoretical prediction $-2r=1/8$.   
}

\kword{classical XY model, Kosterlitz-Thouless transition, 
Monte Carlo simulation, cluster algorithm, finite-size scaling, 
parallel computing, multiple GPU calculation}

\begin{document}

\maketitle



The two-dimensional (2D) classical XY model (planar rotator model) 
shows a unique phase transition, the Kosterlitz-Thouless (KT) 
transition \cite{KT,Kosterlitz}. 
It does not have true long-range order, but the correlation function 
decays as a power of the distance at all the temperatures 
below the KT transition point. 
Compared to the algebraic divergence of the correlation length $\xi(T)$
for the second-order phase transition, 
the correlation length diverges much faster as 
\begin{equation}
  \xi(T) \propto \exp(c/\sqrt{t}) = A \exp(c/\sqrt{t})
\label{corr_length}
\end{equation}
with $t=(T-T_{\rm KT})/T_{\rm KT}$
for the KT transition.  Thus, numerical studies on large systems 
are needed.  Moreover, the difficulties in Monte Carlo simulations 
come from logarithmic corrections that are predicted to be present 
\cite{Kosterlitz,Janke97}.  The magnetic susceptibility scales as 
\begin{equation}
  \chi \propto L^{2-\eta} \ (\ln L)^{-2r}
\label{log_correction}
\end{equation}
with $\eta = 1/4$, and the theoretical prediction of $-2r$ 
is 1/8 \cite{Kosterlitz}. 
There were discrepancies in the estimate of $T_{\rm KT}$ of
two most precise results \cite{Olsson,Hasenbusch1997}; 
to resolve this discrepancy Hasenbusch \cite{Hasenbusch2005} made extensive 
Monte Carlo simulations on lattices up to the linear system size $L=2048$.
There has been a puzzle in the estimate of the exponent to describe 
the logarithmic correction, $-2r$.  The history of the estimates 
of the exponent $-2r$ is listed in the paper by Kenna \cite{Kenna}. 
We note that the critical temperature and the logarithmic-correction exponent 
were also calculated by the high-temperature expansion of 
the magnetic susceptibility to orders $\beta^{33}$ for the classical 
XY model \cite{Arisue}.


The Monte Carlo simulation has served as a standard numerical tool 
in the field of many body problems in physics. 
However, the Metropolis-type single-spin-flip algorithm \cite{metro53} 
often suffers from the problem of slow dynamics or critical slowing down. 
To overcome this difficulty, a multi-cluster flip algorithm was proposed 
by Swendsen and Wang (SW) \cite{sw87}.  
Wolff \cite{wolff89} proposed another type of cluster algorithm, 
that is, a single-cluster algorithm. 


Tomita and Okabe \cite{PCC,tomita2002} developed a cluster algorithm,
which is called the probability-changing cluster (PCC) algorithm,
of locating the critical point automatically.  
It is an extension of the SW algorithm \cite{sw87}, 
but one changes the probability of cluster update 
(essentially, the temperature) during the Monte Carlo process.  
In the original paper \cite{PCC}, Tomita and Okabe investigated 
the second-order phase transition of discrete spin models, 
such as the $q$-state Potts model, but 
the PCC algorithm was extended to the problem of the vector order 
parameter \cite{tomita2002a} with the use of Wolff's embedded 
cluster formalism \cite{wolff89}. 


High performance computing triggers advances in science. 
The use of general purpose computing on graphics processing 
unit (GPU) is a recent hot topic in computer science. 
Drastic reduction of processing times can be realized in 
scientific computations. 
Using the common unified device architecture (CUDA) released by NVIDIA, 
it is now easy to implement algorithms on GPU 
using standard C or C++ language with CUDA specific extension. 
A parallel computing is performed using many threads in a single GPU. 
Moreover, larger-scale problems beyond the capacity of video memory 
on a single GPU can be accommodated by multiple GPUs.  
Multiple GPU computing requires GPU-level parallelization. 
A large-scale open science supercomputer TSUBAME 2.0 
is available at the Tokyo Institute of Technology. 
TSUBAME 2.0 consists of 4224 NVIDIA Tesla M2050 GPUs as a total, 
and the theoretical peak performance reaches 2.4 PFLOPS. 


A successful application of GPU computing to the Metropolis-type 
single-spin-flip algorithm was proposed by Preis {\it et al.} 
\cite{preis09,preis11}.  
The GPU-based calculation of the multispin 
coding of the Monte Carlo simulation 
was reported \cite{block10}, 
where the multiple GPU calculation was argued. 
High performance computing using GPUs 
is highly desirable for Monte Carlo simulations 
with cluster flip algorithms.  Recently, some attempts have been 
reported along this line on a single GPU.  
Weigel \cite{weigel11} has studied parallelization of 
cluster labeling and cluster update algorithms for calculations 
with CUDA. 
He realized the SW multi-cluster algorithm 
by using the combination of self-labeling algorithm and 
label relaxation algorithm or hierarchical sewing algorithm. 
The present authors \cite{komura12} have proposed the GPU 
calculation for the SW multi-cluster algorithm 
by using the two connected component labeling algorithms, 
the algorithm by Hawick {\it et al.} \cite{Hawick_labeling} 
and that by Kalentev {\it et al.} \cite{Kalentev}, 
for the assignment of clusters. 
The computational speed for the 2D $q=2$ Potts model
on NVIDIA GeForce GTX580 was 12.4 times as fast as
that on a current CPU core, Intel Xeon CPU W3680.  
%
More recently, we have extended the GPU-based calculation 
on a single GPU to multiple GPU computation \cite{komura12_multiGPU}. 
To deal with multiple GPUs, we use the message passing interface (MPI) 
library for communication.  We employ a two-stage process 
of cluster labeling; that is, the cluster labeling within each GPU 
and the inter-GPU labeling. 
We have tested the performance for the 2D $q$-state Potts model.  
By using 256 GPUs we have treated systems up to $L=65536$. 

In this paper, we study the 2D classical XY model by the use of 
large scale computations with multiple GPUs on the system of TSUBAME 2.0.  
We use a generalized version of the PCC algorithm based on 
the helicity modulus. 
We locate the KT temperature with the self-adapted approach of 
the PCC algorithm.  We pay attention to the 
logarithmic corrections of the susceptibility;  we study 
the exponent $-2r$ to specify the logarithmic correction. 


The Hamiltonian of the classical XY model is given by
\begin{equation}\label{Hc}
 \mathcal{H} = -J \sum_{<i,j>} \vec s_i \cdot \vec s_j
   = - J \sum_{<i,j>} \cos(\theta_i - \theta_j),
\end{equation}
where $\vec s_i$ is a unit vector with two real components. 
The sum is over nearest-neighbor sites of square lattice ($L \times L$) 
with periodic boundary conditions. 


The helicity modulus gives the reaction of the system under a torsion. 
The Kosterlitz renormalization-group equations lead to 
the universal jump of the helicity modulus \cite{Kosterlitz,Nelson}; 
that is, from the value $(2/\pi) T_{\rm KT}$ to 0 
at $T_{\rm KT}$ in the thermodynamic limit. 
The helicity modulus $\Upsilon$ can be expressed as \cite{Teitel,Weber}
\begin{equation}
 \Upsilon = \langle e \rangle - \frac{J L^2}{T} \langle s^2 \rangle , 
\end{equation}
where
\begin{eqnarray}
 e &=& \frac{1}{L^2} \sum_{<ij>_x} \cos(\theta_i - \theta_j), \\
 s &=& \frac{1}{L^2} \sum_{<ij>_x} \sin(\theta_i - \theta_j),
\end{eqnarray}
and the sum is over all links in one direction.
We note that the definition of $\Upsilon$ in Ref.~\citen{Hasenbusch2005} 
includes $\beta$; that is, $\Upsilon_{\rm Hasenbusch}$ = 
$\beta \Upsilon$. 
Hasenbusch \cite{Hasenbusch2005} calculated the tiny correction 
to the critical value of $\beta \Upsilon$ at $T_{KT}$ 
due to winding field for periodic boundary conditions; the calculated 
critical value is 0.636508, which differs from $2/\pi = 0.636620$ by 0.02\%.


In the original formulation of the PCC algorithm \cite{PCC}, 
one increases or decreases temperature, depending on the observation 
whether clusters are percolating or not. 
The critical temperature is determined by using the finite-size 
scaling (FSS) property of the existence probability of percolation. 
Tomita and Okabe also presented a generalized scheme of 
the PCC algorithm \cite{tomita2002b}
based on the FSS property of the ratio of correlation functions 
with different distances.  
In the present paper, we use the helicity modulus 
as a basis of the PCC algorithm. 
The reason to use the helicity modulus is that for the multiple 
GPU computation there are difficulties in the check of percolation 
and the calculation of correlation function with long distance. 
These difficulties are due to distributed memories 
in the multiple GPU system.   
The actual procedure for the change of temperature is as follows: 
If the helicity modulus $\beta \Upsilon$ is smaller (larger) than 
0.636508, we increase (decrease) $\beta$ by $\Delta \beta \ (>0)$.


We have made simulations for the classical XY model 
on the square lattice with the system sizes 
$L$ = 64, 128, 256, 512, 1024, 2048, 4096, 8192, 16384, 32768, 
and 65536.  Actually, we use a CPU, Xeon W3680, for $L$ up to 256, 
and a single GPU, NVIDIA GeForce GTX580, for $512 \le L \le 4096$.  
We use a multiple GPU system, TSUBAME 2.0, 
for the system $L \ge 8192$.  For multiple GPUs, we assign 
the sub-lattice of size $4096 \times 4096$ for each GPU.  
That is, the systems with $L$ = 8192, 16384, 32768, and 65536 
are realized by 4, 16, 64, and 256 GPUs.

\begin{figure}
\begin{center}
\includegraphics[width=\figwidth\linewidth]{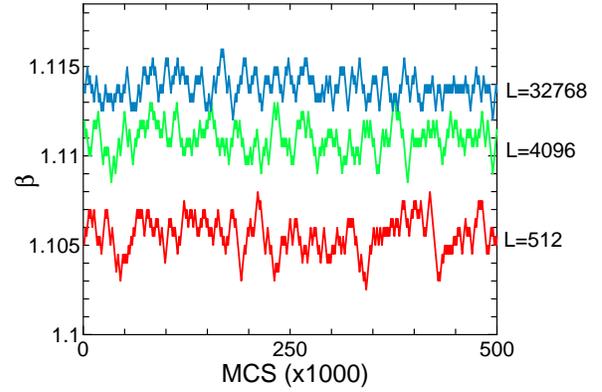}
\caption{
(Color online) 
Time evolution of $\beta$ with the PCC algorithm for the 2D 
classical XY model.  The system sizes are 512, 4096, and 32768.
}
\label{fig:fig1}
\end{center}
\end{figure}

We discard 60000 Monte Carlo steps (MCS) before making a measurement. 
To measure the helicity modulus we take 1000 MCS. 
We choose $\Delta \beta$ as 0.0005.  Throughout this paper 
we use the unit of $J=1$. 
First, we plot the time evolution of $\beta$ in Fig.~\ref{fig:fig1}. 
The time steps are given in units of 1000 MCS.  For illustration, 
the data for $L$ = 512, 4096, and 32768 are shown. 
From Fig.~\ref{fig:fig1}, we see that the temperature is oscillating 
around the average value. 
The width of fluctuation becomes smaller as the system size increases 
because of the effect of self-averaging.  

We take an average of inverse temperature over long steps; 
we denote it by $\beta_{\rm KT}(L)$ for each size. 
We tabulate $\beta_{\rm KT}(L)$ in Table \ref{table1}. 
The numbers of measured steps in units of $10^3$ MCS for each size 
are also given in Table \ref{table1}.  The numbers of measured MCS 
range from $4 \times 10^5$ to $4 \times 10^6$ depending on $L$. 
The uncertainties of measured values, denoted in the parentheses, 
are estimated by the short-time average. 

\begin{table}
\begin{center}
\caption{Estimates of $\beta_{\rm KT}(L)$, $\langle m^2 \rangle _L$ and 
$\xi_{\rm 2nd}(L)/L$ 
for the 2D classical XY model using the PCC algorithm. 
The numbers of measured steps in units of $10^3$ MCS are given.}
\label{table1}
\vspace{2mm}
\begin{tabular}{rrrrr}
\hline 
$L$  \quad & $\beta_{\rm KT}(L)$ \quad\quad & $\langle m^2 \rangle _L$ \quad\quad & 
$\xi_{\rm 2nd}(L)/L$ \quad\quad & $10^3$MCS \\
\hline
64    & \ 1.09167(25)  & \ 0.34924(29)  & \ 0.7464(11) & \ 4000 \\
128   & \ 1.09788(32)  & \ 0.29695(36)  & \ 0.7466(15) & \ 4000 \\
256   & \ 1.10224(21)  & \ 0.25245(28)  & \ 0.7482(09) & \ 4000 \\
512   & \ 1.10530(26)  & \ 0.21427(27)  & \ 0.7482(14) & \ 4000 \\
1024  & \ 1.10768(19)  & \ 0.18195(16)  & \ 0.7488(11) & \ 4000 \\
2048  & \ 1.10966(20)  & \ 0.15443(26)  & \ 0.7491(14) & \ 4000 \\
4096  & \ 1.11097(08)  & \ 0.13085(15)  & \ 0.7495(12) & \ 4000 \\
8192  & \ 1.11211(27)  & \ 0.11086(32)  & \ 0.7500(21) & \ 2000 \\
16384 & \ 1.11314(20)  & \ 0.09395(25)  & \ 0.7492(25) & \ 1000 \\
32768 & \ 1.11389(21)  & \ 0.07955(22)  & \ 0.7499(22) & \  500 \\
65536 & \ 1.11457(14)  & \ 0.06737(14)  & \ 0.7504(16) & \  400 \\
\hline 
\end{tabular}
\end{center}
\end{table}

\begin{figure}
\begin{center}
\includegraphics[width=\figwidth\linewidth]{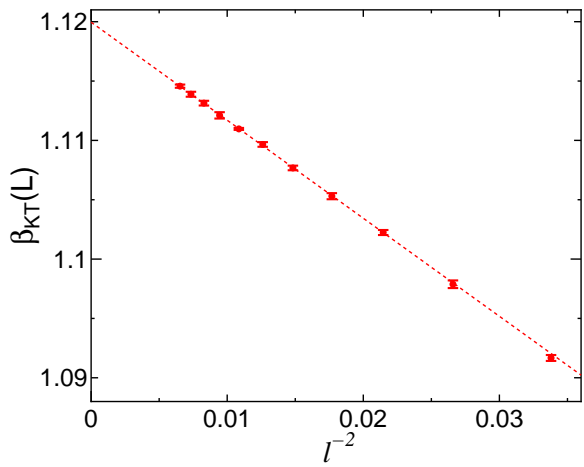}
\caption{
(Color online) 
Plot of $\beta_{\rm KT}(L)$ of the 2D classical XY model 
for $L$ = 64, 128, 256, 512, 1024, 2048, 4096, 8192, 16536, 32768, and 65536, 
where $l = \ln bL$.  The best fitted line, $1.11996 - 0.826 \times l^{-2}$, 
is shown with $b=3.6$. 
}
\label{fig:fig2}
\end{center}
\end{figure}

Then, we consider the size dependence of $\beta_{\rm KT}(L)$.
We use the FSS analysis based on the KT form of 
the correlation length, Eq.~(\ref{corr_length}).
Using the PCC algorithm, we obtain the inverse temperature 
$\beta_{\rm KT}(L)$ such that 
the helicity modulus $\beta \Upsilon$ is $0.636508$. 
It means that $\xi/L (=a)$ is constant for each size.  
Then, using the FSS form of $\Upsilon$, that is, 
$\Upsilon=\Upsilon(\xi/L)$, we have the relation 
\begin{equation}
 \beta_{\rm KT}(L) = \beta_{\rm KT} - \frac{c^2 \beta_{\rm KT}}{(\ln bL)^2}, 
\label{beta_KT}
\end{equation}
where $b=a/A$. 
We plot $\beta_{\rm KT}(L)$ as a function of $l^{-2}$ with 
$l = \ln bL$ for the best fitted parameters in Fig.~\ref{fig:fig2}.  
The error bars, which are very small, are shown there. 
We try several estimates; that is, the minimum size $L$
for the check of $\chi^2$ values is taken as 64, 128, 256, and 512. 
From smaller $\chi^2$ values, we actually use the data of 
the minimum size $L$ being 128 and 256. 
Our estimate of $\beta_{\rm KT}$ is 1.11996(6); the numbers 
in the parentheses denote the uncertainty in the last digits. 
This value is consistent with the estimates of recent studies; 
1.1199(1) by the Monte Carlo simulation \cite{Hasenbusch2005}, 
and 1.1200(1) by the high-temperature expansion \cite{Arisue}. 

\begin{figure}
\begin{center}
\includegraphics[width=\figwidth\linewidth]{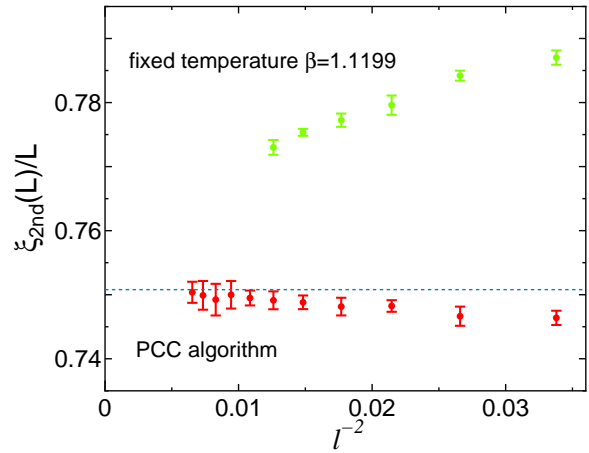}
\caption{
(Color online) 
Plot of $\xi_{\rm 2nd}(L)/L$ at $T_{\rm KT}(L)$ 
for $L$ = 64, 128, 256, 512, 1024, 2048, 4096, 8192, 16536, 32768, and 65536. 
The horizontal axis is the same as that of Fig.~2.  
The data with the PCC algorithm are compared with those 
at the fixed temperature $\beta=1.1199$. 
The calculated value for $L=\infty$ by Hasenbusch \cite{Hasenbusch2005}, 
$0.75069 \cdots$, is given by a dotted line for convenience.
}
\label{fig:fig3}
\end{center}
\end{figure}

Next consider the correlation length of second moment $\xi_{\rm 2nd}(L)$; 
\begin{equation}
  \xi_{\rm 2nd}(L) = \frac{[\chi(0)/\chi(2\pi/L)-1]^{1/2}}{2\sin(\pi/L)} 
\end{equation}
with
\begin{equation}
  \chi(k) = \frac{1}{N}\left|\left\langle \sum_r \vec s(r) e^{ikr}
            \right\rangle \right|^2 .
\end{equation}
In Ref.~\citen{Hasenbusch2005}, $\xi_{\rm 2nd}(L)/L$ is calculated 
as 0.7506912 for $L \to \infty$ at the KT temperature $T_{\rm KT}$. 
We tabulate $\xi_{\rm 2nd}(L)/L$ at $T_{\rm KT}(L)$ in Table \ref{table1}, 
and plot them in Fig.~\ref{fig:fig3}, where the horizontal axis 
is chosen as the same as Fig.~\ref{fig:fig2}.  
We see that $\xi_{\rm 2nd}(L)/L$ approaches $0.750\cdots$ rapidly 
even for small $L$.  We also make simulations at $\beta=1.1199$ 
to confirm the consistency with the calculation 
by Hasenbusch \cite{Hasenbusch2005}. 
For comparison we plot $\xi_{\rm 2nd}(L)/L$ with fixing 
$\beta=1.1199$ in Fig.~\ref{fig:fig3}, which approaches $0.750\cdots$ slowly 
with the increase of size. 

\begin{figure}
\begin{center}
\includegraphics[width=\figwidth\linewidth]{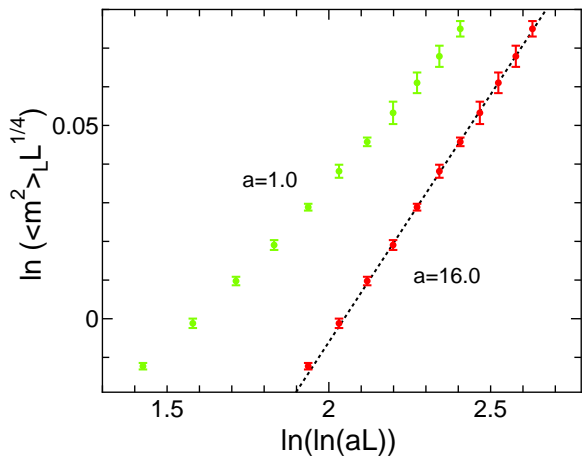}
\caption{
(Color online) 
Plot of $\ln(\langle m^2 \rangle _L L^{1/4})$ as a function of $\ln(\ln(aL))$ 
at $T_{\rm KT}(L)$ for $L$ = 64, 128, 256, 512, 1024, 2048, 4096, 
8192, 16536, 32768, and 65536.  The best fitted line with $a=16.0$, 
$-0.262 + 0.128 \times \ln(\ln(aL))$, is given.  
We also plot the case of $a=1.0$ for comparison
}
\label{fig:fig4}
\end{center}
\end{figure}

We now turn to the logarithmic-correction exponent $-2r$, 
given in Eq.~(\ref{log_correction}). 
It comes from the multiplicative logarithmic corrections 
in terms of correlation length, such that
\begin{equation}
 \langle m^2 \rangle _L = \chi/L^2 \propto \xi^{-\eta} (\ln \xi)^{-2r}.
\end{equation}
Since we measure $\langle m^2 \rangle _L$ at the temperature $T_{\rm KT}(L)$ 
with $\xi/L=a$ for each size, we have 
\begin{equation}
 \langle m^2 \rangle _L \, L^{1/4} \propto (\ln aL)^{-2r}.
\label{m2_2r}
\end{equation}
The measured data for $\langle m^2 \rangle _L$ are tabulated in Table \ref{table1}. 
If we take logarithm of both sides of Eq.~(\ref{m2_2r}), we have 
\begin{equation}
 \ln \Big(\langle m^2 \rangle _L \, L^{1/4} \Big) = {\rm const} - 2r \ln(\ln aL).
\end{equation}
We plot $\ln(\langle m^2 \rangle _L L^{1/4})$ as a function of $\ln(\ln aL)$ 
with the best fitted parameter in Fig.~\ref{fig:fig4}, 
where $a$ is chosen as 16.0. 
Checking the $\chi^2$ values, we determine $-2r$ as 0.128(8). 
In Fig.~\ref{fig:fig4}, we also give a plot of $\ln(\langle m^2 \rangle _L L^{1/4})$ 
as a function of $\ln(\ln L)$; that is, $a=1$.  
If we use only the data of $a=1$ for smaller lattices (say, $64 \le L \le 2048$),
the best fitted slope, which indicates $-2r$, in Fig.~\ref{fig:fig4} 
becomes 0.07; whereas the slope becomes 0.11 when using only the data 
for larger lattices (say, $2048 \le L \le 65536$). 
Thus, for large enough $L$, up to 65536, we can say that the slope, $-2r$, 
approaches the theoretical value 1/8 (=0.125) irrespective of the choice of $a$.

We observe that the slope, which indicates $-2r$, is smaller for small $L$ 
if we do not consider $a$.  For large enough $L$, up to 65536, 
we clearly see that the slope, $-2r$, approaches the theoretical value 
1/8 irrespective of the choice of $a$. 
We may conclude that the puzzle has been finally solved. 

We note that the thermal average of the physical quantities, such 
as $\langle m^2 \rangle _L$, with the PCC algorithm is consistent with 
the $T$-fix calculation within the error bars. 
In the PCC algorithm, $T$ is changing, but the thermal average of 
physical quantity coincides with the $T$-fix calculation 
because the fluctuation width is small enough.
 
We make a comment on the computational time.  
The GPU computation on a single NVIDIA GeForce GTX580 
for the $4096 \times 4096$ system 
takes about 14 hours, including the spin-flip and 
the measurement of the helicity modulus, the susceptibility, 
and the second-moment correlation length for 500,000 MCS. 
When we use multiple GPUs, the computational time for each GPU 
slightly increases with the increase of the number of GPUs 
because of the time for communication.  
Thus, the system of the size $65536 \times 65536$ 
can be treated by using 256 NVIDIA Tesla M2050 GPUs in a day.   

To summarize, we have studied the 2D classical XY model by a large-size 
cluster-update Monte Carlo simulation, and have determined 
$T_{\rm KT}$ using the PCC algorithm based on helicity modulus 
up to the size $L=65536$.  The estimated inverse KT temperature is 
1.11996(6). 
We have shown that $\xi_{\rm 2nd}(L)/L$ approaches $0.750\cdots$ rapidly 
with the increase of $L$ at the temperature where the helicity modulus 
is $0.636508 \times T$.
The logarithmic-correction exponent $-2r$ 
is estimated as 1/8 of the theoretical value with no assumption. 

We have again confirmed that the self-adapted approach of the PCC algorithm 
is efficient to locate the critical temperature, not only for the second-order 
transition but also for the KT transition. 
We finally emphasize that GPU computation is very effective 
when a large-size calculation is needed, for example, 
for systems where logarithmic behavior is essential.

\begin{acknowledgments}
We thank Yusuke Tomita, Takayuki Aoki, and Wolfhard Janke for valuable discussions. 
The computation of this work has been done using 
TSUBAME 2.0 at the Tokyo Institute of Technology. 
This work was supported by a Grant-in-Aid for Scientific Research 
from the Japan Society for the Promotion of Science. 
\end{acknowledgments}

\end{document}